# Vortex oscillations around a hemisphere-cylinder body with a high fineness ratio

Bao-Feng Ma[a)] and Shuo-Lin Yin

*Ministry-of-Education Key Laboratory of Fluid Mechanics, Beihang University, Beijing 100083, China*

In our previous studies (*Ma and Liu, Phy Fluids, 2014; Ma, Huang, and Liu, Chin J Aeronaut, 2014*), the unsteady vortex flows around a pointed-ogive slender body were studied by means of numerical simulations and experiments, and low-frequency vortex oscillations were found over the forebody as angles of attack (AOAs) more than 65º. In this investigation, the vortex unsteadiness around a hemisphere-cylinder body with a fineness ratio (the ratio of length L to diameter D of a body, L/D) of 24 at AOAs of 10º to 80º was studied using Large Eddy Simulation (LES) and Dynamic Mode Decomposition (DMD). The Reynolds number (Re) based on the cylinder diameter of the body is 22000. The results show that the vortex oscillations still exist over the hemisphere-nose slender body, but the vortex behaviors are much different from the ones over the pointed-nose bodies. The vortex oscillation exists over the forebody at the whole range of AOAs, and occurs even at the AOA of 10º. The oscillation is characterized by alternate oscillations of a forebody leeward vortex pair up and down and in-phase swings from side to side. The vortex shedding can be found at the afterbody as AOAs more than 20º, and the shedding region moves forwards gradually with AOAs increasing, and accordingly the region of vortex oscillations contracts and eventually only exists near the nose as AOAs sufficiently high. The vortex oscillation and shedding all induce fluctuating side forces along the body, but the ones from vortex oscillations are larger. The frequencies of vortex oscillations are similar to the ones of vortex shedding at the AOAs of 10º-40º with *St*=0.085-0.12, in which the flow fields over the afterbody are dominated by vortex shedding and forebody-vortex wakes together; while at AOAs of 50º-80º, the frequencies along the body are apparently divided into two regimes in which the vortex oscillations over the forebody have the frequencies of 0.053-0.064, and the vortex shedding over the afterbody has the frequencies of 0.16-0.2. Additionally, as AOAs beyond 40º, the frequencies are basically proportional to the sine function of AOAs. The simulated frequencies agree well with previous experimental results obtained by hotwire velocity measurements (*Hoang et al., Exp Fluids, 1999*). The oscillatory global modes based on the sectional flow snapshots were obtained through DMD analysis. The results show that except the mean flows, the most energetic modes correspond to the vortex oscillation at the forebody and to the vortex shedding at the afterbody respectively, and the frequencies from DMD are identical to the ones of the side forces obtained by fast Fourier transform. In addition, both the time-averaged side forces and vorticity fields show that the mean flow fields for the vortex oscillations are symmetric, and no apparent asymmetry exists. Therefore, the vortex pair over the forebody oscillates around a symmetric mean flow field.

## I. INTRODUCTION

Slender bodies have extensive applications in industries, particularly in aerospace fields. Many studies have been carried out to explore the vortex flow behaviors around axisymmetric slender bodies at angles of attack (AOAs) and associated aerodynamic characteristics. Depending on the nose shapes for slender bodies, the wake flows past a slender body exhibit various interesting phenomena.

---

[a)] Corresponding email: bf-ma@buaa.edu.cn

For the slender bodies with pointed or slightly blunt noses, previous research indicated that the flows are typically divided into four regimes[1, 2] from low to high AOAs: attached flow, steady symmetric vortices, steady asymmetric vortices, and unsteady vortices. In the four flow regimes, the asymmetric steady vortices have been paid much attention in past decades[1-12]. The research has revealed that the vortex asymmetry primarily arises from imperfections on the nose-tip and even natural imperfections from machining tolerance can trigger the development of the asymmetric vortex flows. The asymmetric vortices can induce large time-averaged side forces which are sometimes even beyond normal forces (the side force is the component of aerodynamic force normal to the plane spanned by the vector of freestream velocity and the symmetry axis of a body). The sensitivity of asymmetric vortices over slender bodies to nose-tip imperfections has been exploited for vortex control [4, 6-7, 9-10]. The vortices will become unsteady at sufficiently high AOAs. It can been anticipated that the flow patterns over downstream afterbodies will transition to Kármán vortex shedding if the cylindrical afterbody is sufficiently long[13, 14]. Besides the vortex shedding, however, other unsteady flow patterns also exist [15]. Degani et al [16] examined unsteady characteristics of vortices over slender bodies in experiments, revealing three types of unsteady phenomena by low-frequency Kármán vortex shedding, high-frequency shear-layer unsteadiness, and vortex interaction at moderate frequencies. Among these unsteady phenomena, low-frequency vortex shedding has the largest fluctuating amplitude. The experiments by Zilliac et al [2] using a smoke visualization qualitatively showed that the flows could be divided into three patterns along the body axis at the AOAs of more than 65 deg: a pair of stationary vortices over a forebody, oblique vortex shedding over the cylinder part adjacent to the forebody, and parallel vortex shedding around the afterbody. However, Ma et al.'s results in experiments [17] and numerical simulations [18] indicated that the forebody vortex pair is not stationary beyond 65º AOA, but oscillates around a time-averaged asymmetric orientation with much lower frequencies than vortex shedding. Therefore, the leeward vortices around sharp-nosed slender bodies are essentially steady at most of the AOAs, including steady symmetric and asymmetric vortices, while large-scaled vortex motions exist only at sufficiently high AOAs (more than 65 º typically ).

Nevertheless, for the slender bodies with a hemispherical nose, the vortex flows become very different from the cases with pointed noses. Previous studies [19-20] showed that the time-averaged vortex structures over the hemisphere cylinder would vary with increasing AOAs. At lower AOAs of 0º-5º, only nose separation bubble exists near the hemisphere nose due to streamwise separation; at medium AOAs of 5º-30º, a pair of horn vortex occurs on both sides of the nose separation bubble, meanwhile a pair of leeward vortex is also developed downstream along both sides of the body due to cross flow separation; at high AOAs of 30º-50º, the nose separation bubble becomes smaller with increasing AOAs, and eventually merge together with leeward vortices so as to form a single horseshoe vortex system, while the horn vortices disappear; as AOAs more than 50º, no data are available as yet. In addition, the study of Hoang et al. [21] revealed that the horn vortices existing at medium AOAs would become smaller in size with increasing Reynolds number (Re). Past research [6] also showed that the tip sensitivity of the vortex system over slender bodies would be reduced greatly with nose blunting, so the onset of the asymmetric vortices over a hemisphere cylinder will be delayed or suppressed significantly relative to pointed-nose bodies. Experimental studies on time-averaged flow structures indicated that no natural asymmetric vortices were found over hemisphere cylinders[19-21] or a quasi-hemisphere cylinder [22], but asymmetric vortices could be triggered if an artificial disturbance was fixed on the nose [20]. However, recent experimental results about a hemisphere cylinder by force measurements [23] showed that side forces could be obtained at high AOAs, which meant time-averaged asymmetric vortices might exist. Additionally, the results of Jiang et al.[24] by direct numerical simulations showed that an inclined prolate spheroid can produce time-averaged asymmetric vortices even without any artificial disturbance, and the authors attributed this phenomenon to a global instability. Therefore, if the asymmetric vortices exist over a hemisphere cylinder, the underlying mechanisms will be probably similar to the case for the prolate spheroid. Flow unsteadiness over hemisphere cylinders has also been studied in experiments and numerical simulations [25-30], but not so many. Hoang et al.[25] measured flow fluctuations over a hemisphere cylinder at various AOAs using hot wires, in which the Reynolds number based on cylinder diameter was 22000. The study found a type of velocity fluctuations with the frequencies lower than Kármán vortex shedding, and they suggested that the low frequency fluctuations probably come from vortex heaving, but no information on flow patterns was provided. The experiments by Sirangu and Ng [23] using a smoke visualization also showed the vortices around the afterbody of a hemisphere cylinder could change their orientations with time.

Recent years, Direct Numerical Simulation (DNS) and stability analysis in lower Reynolds numbers have been carried out in order to reveal the unsteady flow physics around hemisphere cylinders. Sanmiguel-Rojas et al [26] and Bohorquez et al [27] studied three dimensional (3D) vortex shedding around the rear end of a blunt-nose cylinder at AOA of 0º, and identified a type of global mode of 3D vortex shedding with a non-dimensional frequency of 0.12. Gross et al [28] carried out a DNS to study the flow past a hemisphere-nose axisymmetric body at Re=5000, AOA=10º and 30º, and found that the nose separation bubble around the hemispherical nose could shed downstream at a lower AOA (10º), but the separation bubble shedding was suppressed at a higher AOA (30º). They also found that the unsteadiness of leeward vortices seems to be related to the flow separation around the nose. Le Clainche et al.[29, 30] also conducted a DNS on a hemisphere cylinder at Re=350, 1000 and AOA=20º, in which a type of oscillatory motion for leeward vortices over the hemisphere cylinder was demonstrated. This finding is interesting, because it seems to imply that a type of new global mode might exist over hemisphere cylinders. However, the oscillatory frequencies obtained by Le Clainche et al. [29, 30] are different from the ones in previous experiment [25] due to the much lower Reynolds numbers. The studies of Le Clainche et al. [29, 30] also revealed 3D spatial structures on the separation bubble, horn vortices and leeward vortices over a hemisphere cylinder. Le Clainche et al. [30] suggested that it could not prove that the oscillation of horn/leeward vortices and the wake vortex shedding were independent phenomena because they have similar frequencies.

The purpose of the present study is to confirm the existence of large-scaled vortex oscillations around a hemisphere cylinder, to investigate the evolution and global modes of unsteady flows from low to high AOAs using Large Eddy Simulation (LES) and Dynamic Mode Decomposition (DMD) analysis [31-33]. The Reynolds numbers in simulations are kept the same with the one of Hoang et al. (Re=22000) in order to validate the numerical results. The model is a hemisphere cylinder with a high fineness ratio, in which a longer cylindrical part than the model of Hoang et al. is taken to better simulate Kármán vortex shedding past the afterbody. Numerical simulations have the advantage of avoiding the model vibration encountered in unsteady experiments, which is difficult to remove completely. In order to reveal the vortex oscillations how to influence the aerodynamic characteristics, the time histories and frequency spectra of sectional side-forces are extracted, and the DMD is subsequently utilized to decompose the spatial flow fields by frequencies for finding dominant flow patterns responding for the fluctuation of side-forces. The 3D instantaneous flow fields and the evolution of sectional vorticity patterns with time are also presented.

## II. NUMERICAL METHODS AND MODEL

The numerical simulations were carried out using an unstructured finite volume slover Fluent[TM] 14.5, which has been extensively verified and validated [34]. The computation is based on the full three dimensional incompressible Navier-Stokes equations and Large Eddy Simulation (LES). The dynamic Smagorinsky model[35, 36] is utilized as a subgrid-scaled model. The numerical algorithm is the pressure-based coupled methodology with a Rhie-Chow interpolation scheme[37] to prevent the decoupling of the pressure and velocity fields. A second-order implicit backward-Euler scheme for time stepping was employed, while the advection terms were evaluated using a second order bounded-centered scheme[38, 39] with a low numerical dissipation, and the diffusion terms were calculated using second-order centered scheme. The bounded-centered scheme has successfully been applied to Large Eddy Simulation in previous research[18, 40].

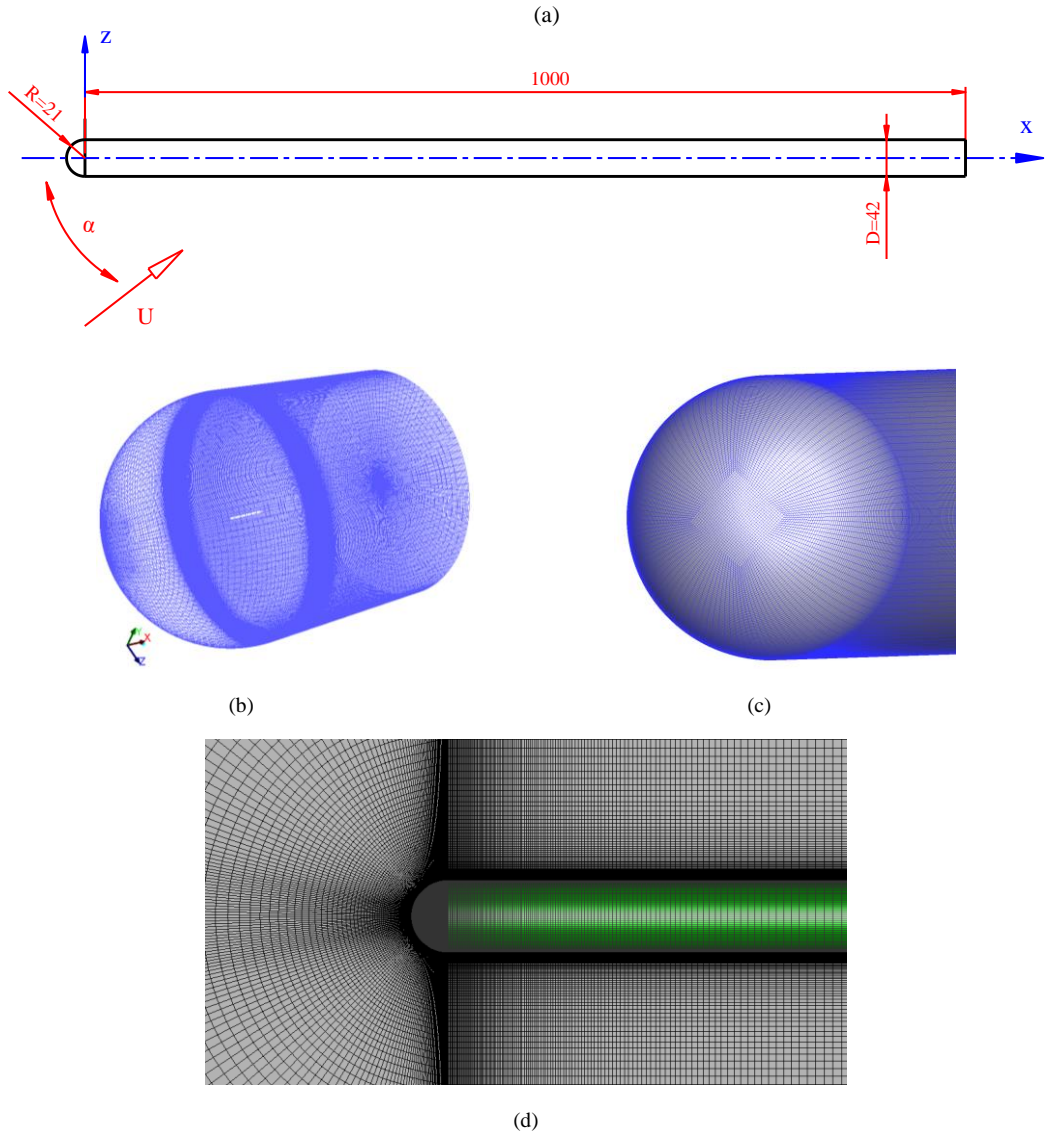

FIG. 1. Numerical model and grids, in mm. (a) computational domain; (c) grids on the hemisphere nose; (d) sectional and wall grids.

The flow fields obtained were analyzed by DMD. The DMD is a good method to find coherent structures in complex flow fields[31], and its mathematical foundation is Koopman operator in dynamical systems[32]. The DMD decomposes the flows by frequencies, therefore enables a comparison between the frequencies of DMD modes and the frequencies of the unsteady aerodynamic force obtained by Fast Fourier Transform (FFT). As a result, the dominant flow structures responsible for the forces can be revealed. The current DMD code is based on the companion matrix algorithm proposed by Schmid[33] and has been verified and validated using analytical synthetic solutions and the flow around a two dimensional circular cylinder. The convergence of the DMD results with sampling snapshots has also been tested in our previous[18] and current studies. The testing all indicated that the 800 snapshots are adequate for DMD.

The computational model is an axisymmetric hemisphere-cylinder slender body, as shown in Figure 1(a). The model has the same diameter with the one in the experiments of Hoang et al.[25], but was designed a longer cylindrical part in order to simulate Kármán vortex shedding past the afterbody. The model of Hoang et al.[25] has a fineness ratio of 8, and the fineness ratio of the present model is extended to 24.3. A cylindrical computational domain with 3.5 times body-length was designed around the x direction of the body, as depicted in Figure 1(b, c, d). The computational domain extended 4.5 times body-length downstream from the rear end of the cylinder and 3.5 times body-length upstream from the nose-tip. A larger computational domain is also tested to exclude the possible impacts of the domain on the results, and the results show that the present domain size is appropriate. The mesh was refined close to the surface of the body in both axial and radial directions to capture flow gradients accurately. The whole outer surface of the computational domain except for the downstream end surface is divided into an upwind surface and downwind surface. A velocity inlet condition was applied at the upwind surface, and a pressure outlet condition is set at the downwind surface and downstream end surface. The wall was set with no slip condition. The grids were created by using the meshing tool Pointwise.

The Reynolds number in the simulation is $2.2 \times 10^4$ based on the cylinder diameter, so the flow is located at the range of subcritical Re regime where boundary layers exhibit laminar separation[3, 8]. Since boundary layers are laminar before separation, the grid density within boundary layers is not necessarily required as much as the case for turbulent boundary layers in LES. The angles of attack are from 10° to 80° with an interval of 10°. Flows were run for a sufficiently long period of time (more than 3 $s$) until regular vortex oscillation and shedding were obtained.

A grid convergence testing was conducted at AOA of 30° with three different grids, as shown in Table I, and the associated grid convergence index (GCI) [41] for estimating the discretization errors was also calculated. The mean drag coefficient ($C_D$) on the body was selected as a global quantity to show the grid convergence, as conducted in previous studies [29, 42]. Meanwhile, since the present investigation was focused on the vortex oscillation, the non-dimensional frequencies ($St$) of the total side-force on the body were also calculated in the grid refinement study and compared with published experimental data [25]. The GCI is defined as GCI $_{j+1, j}$ (%) $=3|(f_{j+1}-f_j)/f_j (r^p-1)| \times 100$ where the $f$ is any integral or field quantities on coarse ($j+1$) and fine ($j$) grids, $r$ is the refinement ratio from fine to coarse mesh sizes, and $p$ is the observed order of accuracy for numerical methods used. The order $p$ can be calculated by the formulas $p=\log (1/R)/\log(r)$ where the convergence ratio $R=(f_{j+1}-f_j)/ (f_{j+2}-f_{j+1})$. Here the $R$ values for drag coefficients and St of the side force all fall within 0<R<1.

It can be seen in Table I that the $C_D$ and $St$ quickly begin to converge as increasing the grid density and the $St$ agrees well with the experimental value. Since the experimental data [25] were acquired by means of hotwire measurements, only the frequencies can be compared. Based on the grid-convergence results, the $8.0 \times 10^6$ grid cells were used for formal computation. The temporal step is 0.0005 *s* (0.0907 in a dimensionless form) for most cases, and a smaller step with 0.00025 *s* (0.0454) was also tested to validate the effect of temporal steps. In the present investigation, the unsteady characteristics of vortex flows over a hemishpere cylinder can be divided into two types by AOAs (see the results in section III): at AOAs of 50º-80º, the vortex oscillation and vortex shedding along the body are similar to the one over pointed-nose slender bodies at high AOAs, but at AOAs of 10º-40º, the behaviors of the vortex oscillation and shedding are different from previous studies. The grid convergence testing and experimental comparison have ever been carried out in our previous simulations for the unsteady vortex flows over a pointed-nose slender body with the AOA of 70º [18], and the results indicated that around eight million grids are sufficient for simulating vortex oscillations. Therefore, here the unsteadiness of vortices at AOAs of 10º-40º was focused on the grid convergence testing, and the case at 30º AOA is a typical one for this type of unsteadiness.

TABLE I. Grid convergence index (CGI) for mean drag coefficients of the whole body and Strouhal number of the total side force at AOA of 30º. A-axial grid, C-circumferential grid, and R-radial grid; $F_D$ is drag, $S$ is the area of a cross section of the cylinder, $\rho$ is airflow density, $U$ is freestream velocity, $D$ is diameter of the cylinder, and $f$ is dimensional frequency

| Total grid number | Time steps for iteration $\Delta t\, U/D$ | Grid distributions around the body (xyz) (From tip to end) | Drag coefficients on the whole body $C_D = F_D/0.5\rho U^2 S$ | CGI ($C_D$) (%) | Strouhal number of the whole side force $St = f\, D/U$ | CGI ($St$) (%) |
|---|---|---|---|---|---|---|
| $2.8 \times 10^6$ | 0.0907 | A170×C80×R130 | 0.19340 | --- | 0.10811 | --- |
| ~ $4.7 \times 10^6$ | 0.0907 | A190×C120×R120 | 0.19662 | 12.0 | 0.11024 | 16.2 |
| ~ $8.0 \times 10^6$ | 0.0907 | A280×C120×R150 | 0.19777 | 4.3 | 0.11109 | 6.4 |
|  | 0.0454 |  | 0.19777 |  | 0.11109 |  |
| Experiment[25] |  |  |  |  | 0.11 |  |

## III. RESULTS

In the section, the sectional side forces are firstly presented in the subsection A to show aerodynamic forces induced by unsteady vortical flows. The associated spatial flow structures are given in the subsection B. Finally, the DMD results on sectional flow fields are provided to further establish the relationship of the fluctuations of side forces with unsteady flow fields in the subsection C.

### A. SECTIONAL SIDE FORCES

The sectional side-force coefficients ($C_y$) along the body axis here is selected as an indicator to show the variations of unsteady aerodynamic forces induced by vortex oscillations and vortex shedding. The $C_y$ is a non-dimensional quantity defined by $C_y = F_y / (0.5\rho v^2 D)$, and the $F_y$ was calculated from the line integral of wall pressure around the circumferential direction at each x/D station ( $F_y = \int_0^{2\pi} pR \sin\theta d\theta$ ; $p$ is wall pressure; $R$ is cylinder radius; $\theta$ is azimuthal angle in a polar coordinate of a cross section of the cylinder). The magnitude of $C_y$ can reveal the level of vortex asymmetry around the body, as commonly utilized in the studies on time-averaged asymmetric vortices over slender bodies[1-12], and here the variation of $C_y$ indicates the evolution of flow asymmetry with time.

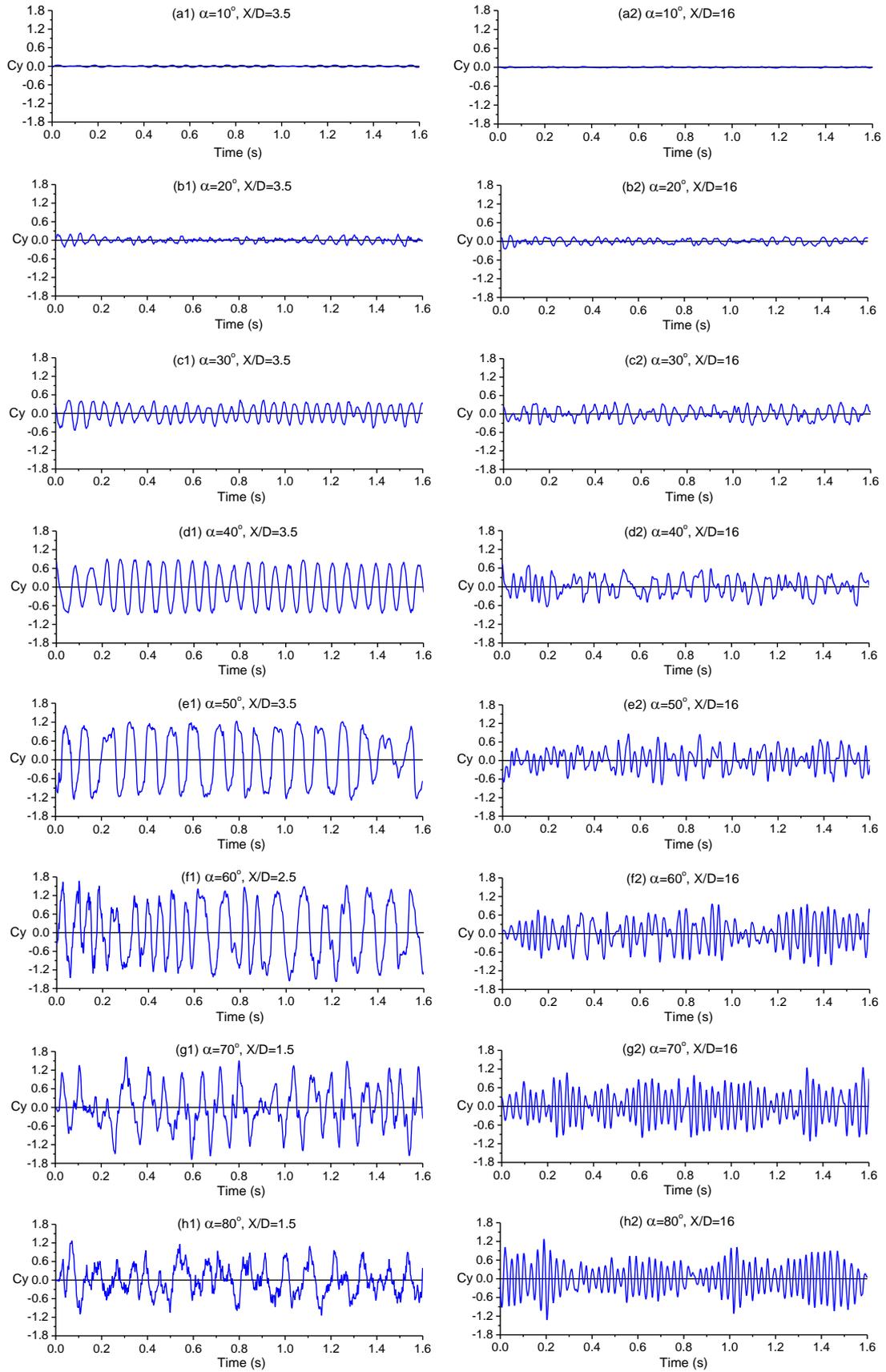

FIG. 2. Time histories of instantaneous sectional side-force coefficients $C_y$ at various AOAs. (a1-h1) Time histories of $C_y$ on the sections of a forebody; (a2-h2) Time histories of $C_y$ at the sections of an afterbody

The time histories of $C_y$ are shown in Figure 2, in which graphs (a1-h1) and (a2-h2) show the $C_y$ fluctuations on the forebody and afterbody, respectively. The side forces start to fluctuate with small magnitude alone the body from the AOA of 10º, and the amplitude of fluctuation roughly increases with increasing AOAs for both the forebody and afterbody with exceptions of the forebody sections beyond 70º AOAs where the side forces decrease. Comparing to the $C_y$ fluctuation at the afterbody, the forebody fluctuations are more regular with larger amplitude between the AOAs of 30º-70º, but the fluctuating amplitude at the fore and afterbody is similar at lower AOAs of 10º-20º. Figure 2 also shows that the mean sectional side forces oscillate almost near a zero value, and actually the time-averaged sectional side forces along the body at various AOAs are approximately zero for all these sections of the body relative to the magnitude of instantaneous $C_y$, which implies that time-averaged flow fields over the body are symmetric. Figure 3 depicts the root-mean-square (*rms*) and frequencies of $C_y$ fluctuation, and the results are presented separately by the range of AOAs in order to more clearly display the trend of variation. Figure 3(a) shows the variation of magnitude at the AOAs of 10º-40º, and it can be seen that the forebody sections (roughly x/D<7) have higher *rms* values, while the *rms* values at the afterbody sections (x/D>7) are relatively small. The frequencies shown in Figure 3(b) indicate that the fluctuation from the front to rear alone the body exhibit similar frequencies with 0.085-0.12. Although the frequencies are not completely the same for the forebody of x/D<9 and afterbody of x/D>9, they are very close. By contrast, the situation at the range of the AOAs of 50-80º are different as shown in Figure 3(c, d). Although the variation of the *rms* are similar to the former where the large fluctuations still are located at the forebody, the frequency characteristics are much different. The distributions of frequencies alone the body are apparently divided into two regimes: one is a low-frequency regime with 0.053-0.064 at the forebody, and another is a high-frequency one with 0.16-0.2. The low-frequency region contracts gradually as the high-frequency region moves forwards with increasing AOAs, as shown in Figure 3(d).

Figure 4 (a) shows the variation of non-dimensional frequencies *St* with increasing AOAs, and another definition on non-dimensional frequencies *St*/*Usinα* is also presented in Figure 4 (b), in which the velocity *Usinα* (α is AOA) is selected as a characteristic velocity. The *rms* basically increases first with increasing AOAs and then decrease. The experimental data obtained by hotwire measurements in the previous study [25] are also shown as compared with the numerical results. The frequencies obtained by the numerical simulation show good agreement with the previous experimental data, except for the cases at less than 20º AOA where the numerical results slightly overestimate the frequencies relative to the experiment. From Figure 4 (a), all sectional frequencies collapse together for each AOA as AOAs less than 40º, showing the frequencies are similar at all sections, which is consistent with Figure 3 (b). In this range of AOAs, the frequencies *St* first slightly increase with increasing AOAs, then decrease. For AOAs more than 40º, the frequencies are apparently divided into two clusters, corresponding to the low frequencies at the forebody and the high frequencies at the afterbody. From Figure 4 (b), it also can be seen that as AOAs beyond 40º, the frequencies are basically unchanged with the exceptions of the transitional region between the vortex oscillation and shedding as well as the region near the rear end of the body, indicating that the frequencies are approximately proportional to the sine function of AOAs at this range of AOAs.

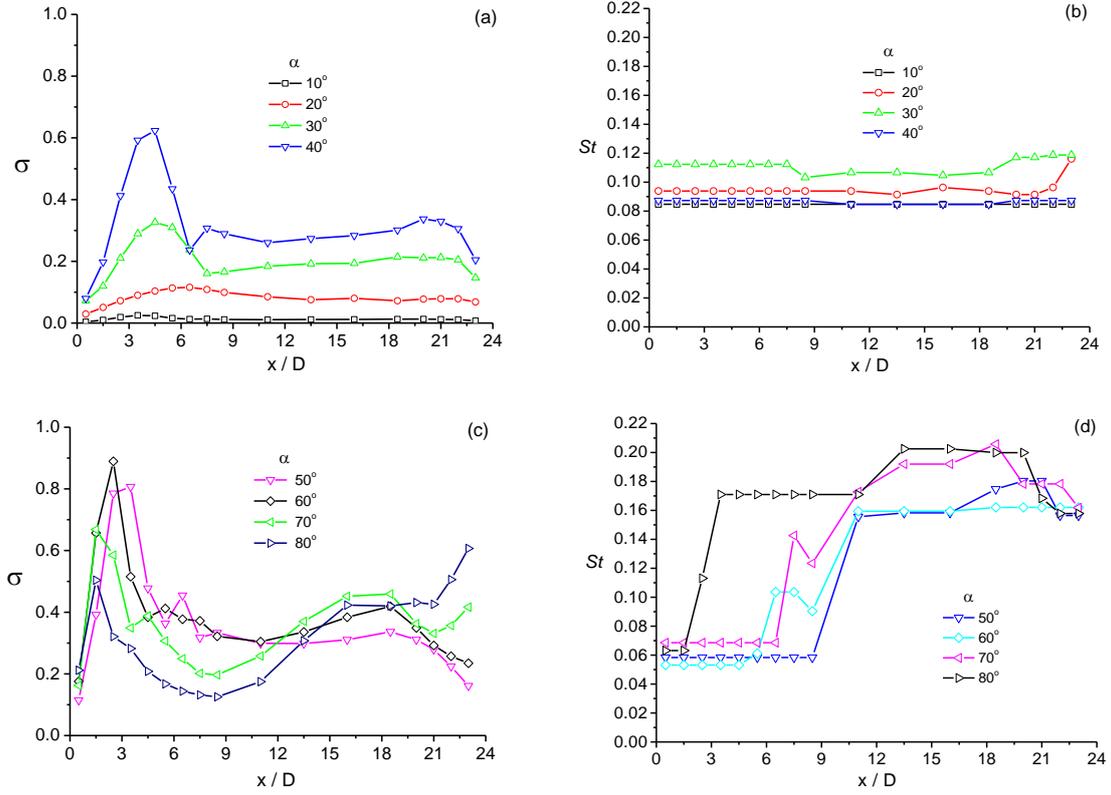

FIG. 3. The magnitudes and non-dimensional frequencies of $C_y$ at various AOAs. (a) Root-mean-square (σ) of $C_y$ at various sections; (b) Non-dimensional frequencies at various sections

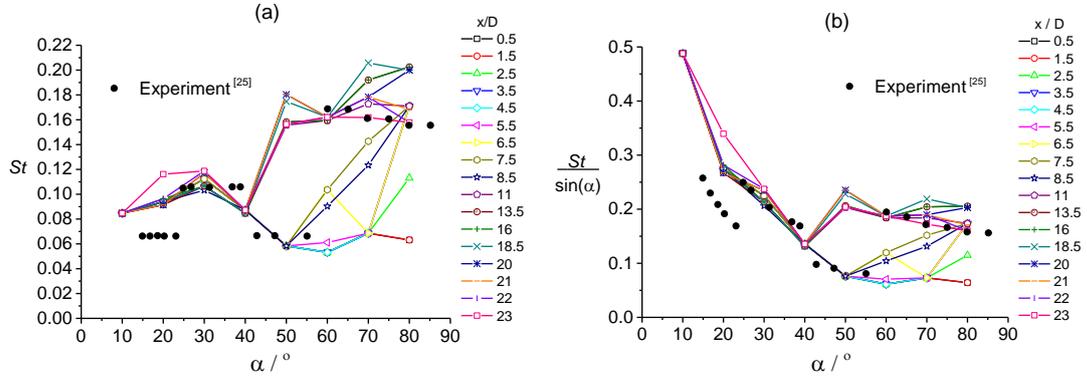

FIG. 4. Variation of non-dimensional frequencies with AOAs. (a) $St$ versus AOAs; (b) $St / \sin(\alpha)$ versus AOAs

## B. SPATIAL FLOW FIELDS

The instantaneous global flow patterns over the hemisphere cylinder at various AOAs are shown in Figure 5, and the vortex regions were identified using the $\lambda_2$-criterion proposed by Jeong & Hussain [43], and the x component of vorticity along the x-axis direction is also presented in color to help distinguishing the left and right vortices and to show their strength. The vortex patterns over the hemisphere cylinder vary with increasing AOAs, and at each AOA, the regions for vortex oscillations and shedding were labeled along the body. The boundaries of the two unsteady vortex regimes were determined by DMD modes of sectional flow fields because the modes for vortex oscillations and shedding are apparently different (see section C). However, the mode types in the adjacent regions of the two flow regimes are difficult to identify probably due to the movement of the boundaries with time, so these regions are termed by transitional regions, denoted by "Tr". The spatial resolutions of the boundaries are dependent on separations between sampling sections (about 1D).

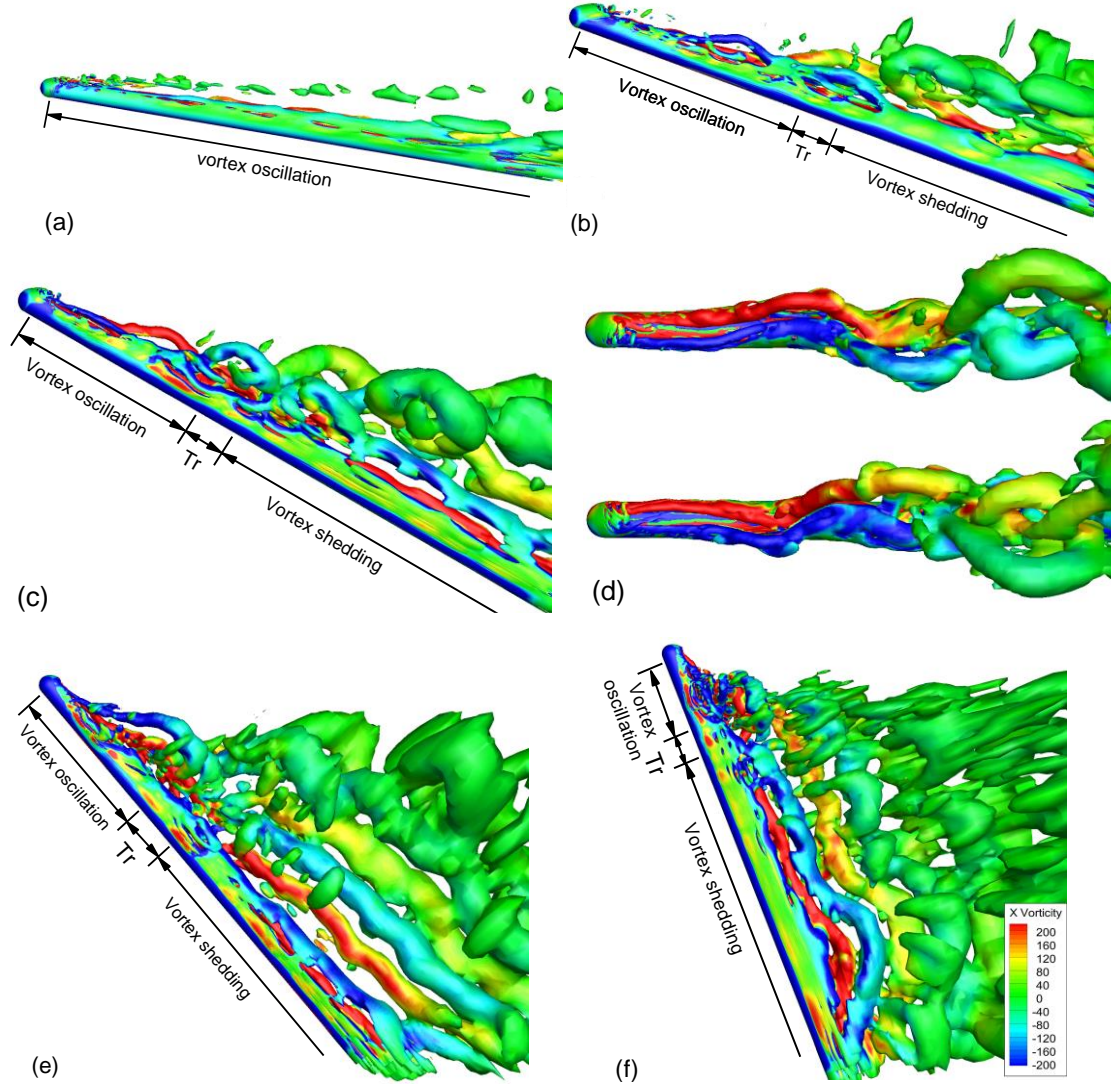

FIG. 5. Vortex structures obtained using $\lambda_2$ method of vortex identification ($\lambda_2$=-0.01; x vorticity denotes magnitude of x component of vorticity, 1/s); Tr denotes the transitional region between two unsteady vortex regimes. (a) α=10º; (b) α=20º; (c) α=30º; (d) α=30º, vortex orientations at two instants, righ swing in phase (upper) and left swing in phase (lower) ; (e) α=50º; (f) α=70º

Figure 5(a) shows the vortex structure of the body at AOA of 10º, where the boundary layers separate from both sides of the body, producing a pair of leeward vortex along the body, and a nose separation bubble and a pair of horn vortices also exist near the hemisphere nose, as revealed in previous studies [19-21, 29-30]. The leeward vortex pair slightly oscillates with a definite dominant frequency so as to induce the fluctuation of side forces, while the bubble and horn vortices seem unstable at this Reynolds number and shed downstream. The shedding of the separation bubble around the nose is gradually weakened with increasing AOAs, and eventually disappears before around 30º AOA. This trend is consistent with the previous study [28] where a hemisphere cylinder with a fineness ratio of 4.81 was investigated numerically, and the separation bubble shedding near the nose was also found, and would disappear with increasing AOAs. At the AOA of 20º, except for the separation bubble and horn vortices near the nose (still shedding, but become weaker), the vortex system along the body can be divided into two regimes: one is leeward vortex oscillations over the forebody, and another is vortex shedding around the afterbody. The forebody leeward vortices break away at somewhere downstream along the body and turn to free vortices, so the separated boundary layers originally feeding the leeward vortices are disrupted. As a result, the disrupted separated boundary layers are reorganized into vortex shedding underneath the free vortices, so the afterbody flow fields are dominated by the combined effects of the downstream wakes of the forebody vortex and the afterbody vortex shedding. The vortex shedding should contribute to the most of the fluctuation of side forces owing to being closer to the wall, while the forebody side forces primarily come from the leeward vortex oscillation. It should be noted that the wakes of the forebody vortices above the vortex shedding are developed downstream with helical vortex tubes, which seems implying the downstream wakes of the forebody vortices have lost stability, like the wake vortex pair after a large airplane [45].

With increasing AOAs, the regions for vortex shedding at the afterbody expand towards the nose, correspondingly the regions for vortex oscillations move forwards, as shown in Figure 5(c, e, f). The forebody vortices not only alternately oscillate up and down, but also swing from side to side. Figure 5(d) illustrates two instantaneous vortex patterns in a top view, where the forebody vortex pair regularly swings to one side in phase, then to the other side. The unsteady side-shift is easy to understand, because a side shift necessarily occurs owing to the mutual induction between vortices as a vortex pair becomes asymmetric in vertical positions or strength (see Figure 6 and 7) [45]. As AOAs are sufficiently high, the vortex shedding dominates the most region of the body, and the forebody vortex pair is eventually limited near the nose, as depicted by Figure 5(e, f).

The instantaneous sectional vorticity fields are shown in Figure 6 and 7 in order to demonstrate the forebody leeward vortex oscillation in the plane normal to the x body axis. The sectional results clearly indicate that the forebody leeward vortices alternately oscillate within nearly one period at both lower AOA (30º) and higher AOA (50º). At the AOA of 30º (Figure 6), the vortex pair initially exhibits an orientation with a higher vortex on the left (Figure 6(a)), and gradually transition to a higher vortex on the right (Figure 4 (b-d)), and eventually the left vortex recovers to the higher position again (Figure 6 (e-f)). The vortex evolution at the AOA of 50º (Figure 7) is similar to the above case. It also can be seen that the vortex pair will be inclined to one side and become a higher vortex as the vortices become asymmetric due to oscillation, which corresponds to the three dimensional vortex evolution in Figure 5(d). Figure 8 shows the time-averaged flow fields for the forebody sections with various AOAs, in which the vortex pair is symmetric. Therefore, the forebody vortex system virtually oscillates around a symmetric mean flow. These results also indicate that the average results are unable to reveal all behaviors of vortices. More importantly, time-averaged symmetric flows are likely caused by statistical averaging so that large-scaled unsteadiness is probably neglected, as revealed by Grandemange et al.[44]. Therefore, average symmetric flow fields are sometimes misleading, because they probably come from either steady symmetric flows or averaging of unsteady flows.

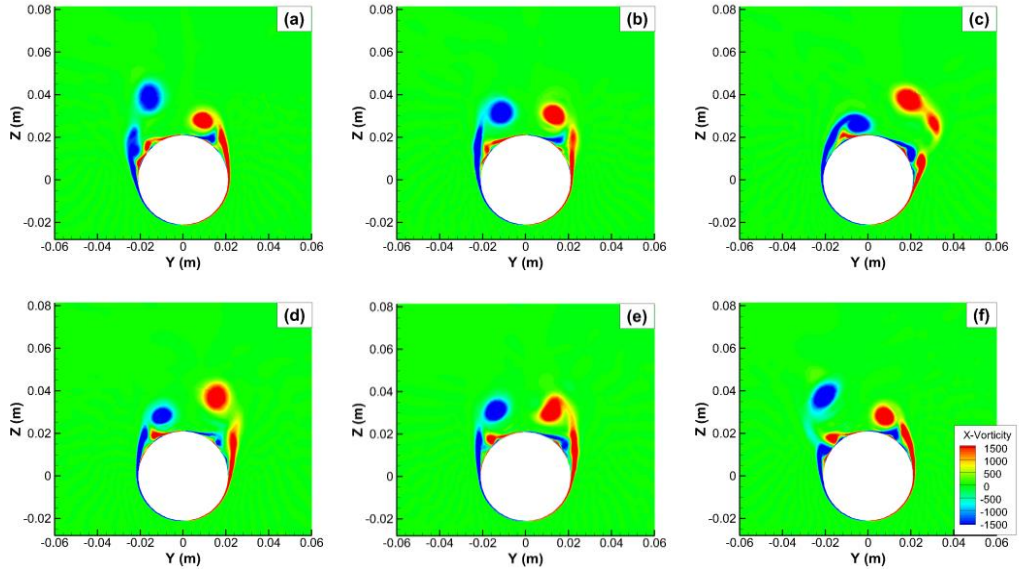

FIG. 6. Vortex evolution at α=30º, and x/D=3.5 within one period, from a rear view; x-vorticity denotes magnitude of x component of vorticity (unit: 1/s). (a) time=$t_0$ s ($t_0$ is a secleted initial value); (b) time= $t_0$+0.007 s; (c) time= $t_0$+0.0125 s; (d) time= $t_0$+0.023 s; (e) time= $t_0$+0.0285 s; (f) time= $t_0$+0.0415 s

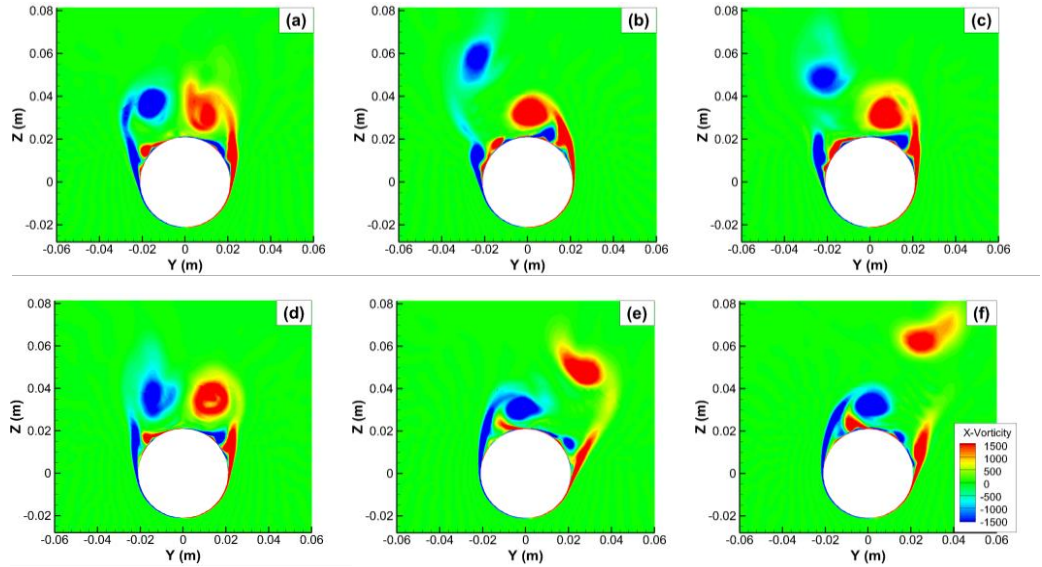

FIG. 7. vortex evolution at α=50º, and x/D=3.5 within one period, from a rear view. (a) time=$t_0$ s ($t_0$ is a secleted initial value); (b) time= $t_0$+0.029 s; (c) time= $t_0$+0.037 s; (d) time= $t_0$+0.046 s; (e) time= $t_0$+0.06 s; (f) time= $t_0$+0.075 s

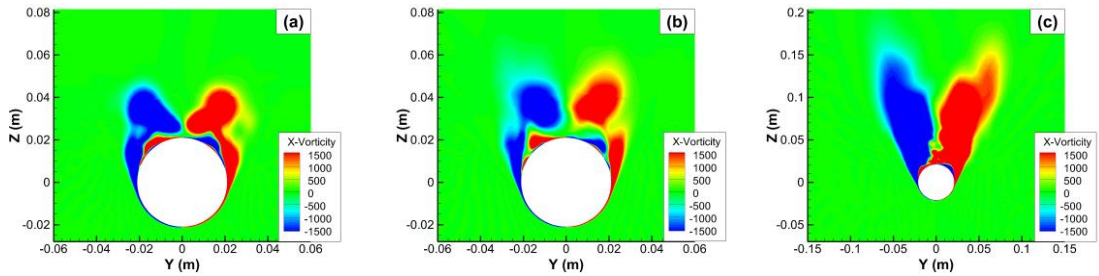

FIG. 8. Time-averaged vorticity patterns at x/D=3.5. (a) α=30º; (b) α=50º; (c) α=70º

## C. DYNAMIC MODE DECOMPOSITION (DMD)

This subsection presents the DMD results of sectional flow fields to further reveal the behaviors of vortex oscillations and to find the oscillatory modes contributing most to the fluctuation of side forces. The spatial flow fields contain more information, so the most energetic flow structures never necessarily produce large aerodynamic forces. The flow structures will have a small effect on the aerodynamic force if they are far away from the wall. Therefore, the DMD is helpful to find the dominant coherent structures responsible for the force, in which the flow modes are decomposed by frequencies which is easy to establish the relationship with the frequencies of the side forces obtained by fast Fourier transform.

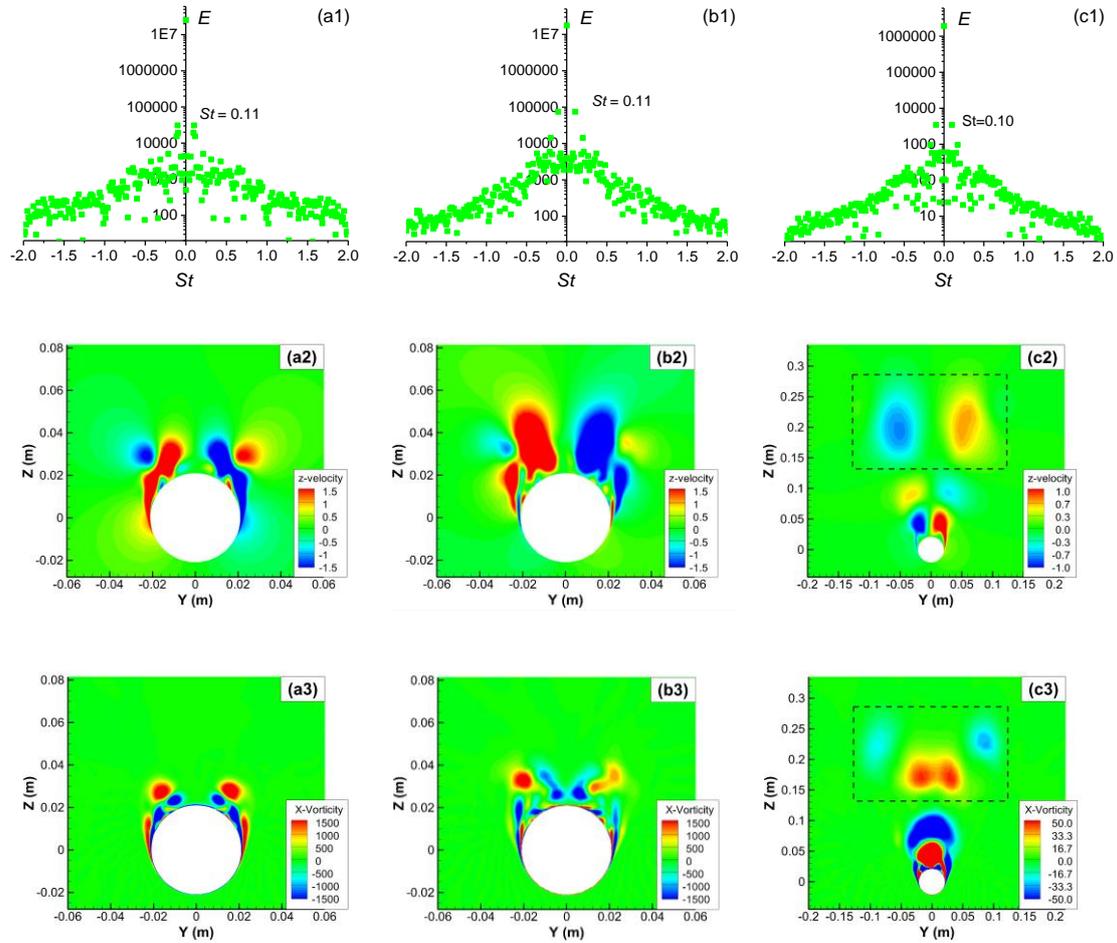

FIG. 9. Dynamic Mode Decomposition (DMD) at AOA=30º (z-velocity is z component velocity, m/s; x-Vorticity is x component of vorticity, 1/s). (a1-a3) x/D=2.5; (b1-b3) x/D=3.5; (c1-c3) x/D=16

Figure 9-11 present the results of DMD for the sectional flow fields at the AOAs of 30º, 50º and 70º. DMD was conducted using 800 snapshots. The energy distributions of various frequencies ($St$) obtained by DMD are shown in (a1, b1, c1) of each graph, and the dominant modes for the dominant frequencies are illustrated by z-component of velocity fields (z-velocity) in (a2, b2, c2) and x-component of vorticity fields (x-Vorticity) in (a3, b3, c3). In the DMD computation, the DMD was first applied to the velocity fields with three components at each section, and then based on the modes of velocity fields obtained, the x-component vortcity fields of the modes can be calculated. Two

forebody sections at each AOA are selected to show the modes of vortex oscillations, and for the AOAs of 30º and 50º, the sections shown are x/D=2.5 and 3.5, while for the AOA of 70º, the sections are presented with x/D=1.5 and 2.5 due to the region of the vortex oscillation move forwards with increasing AOAs; one afterbody section with x/D=16 was chosen for showing the vortex shedding. The modes with maximum energy correspond to the mean flows, which are shown by the data points on the vertical axis. Besides the mean flows, the frequencies for the second energetic modes at each section are identical to the ones of the $C_y$ fluctuaiton as shown in Figure 3-4, which fully demonstrates that the fluctating side forces essentially arise from the most energetic oscillatory modes in the flow fields.

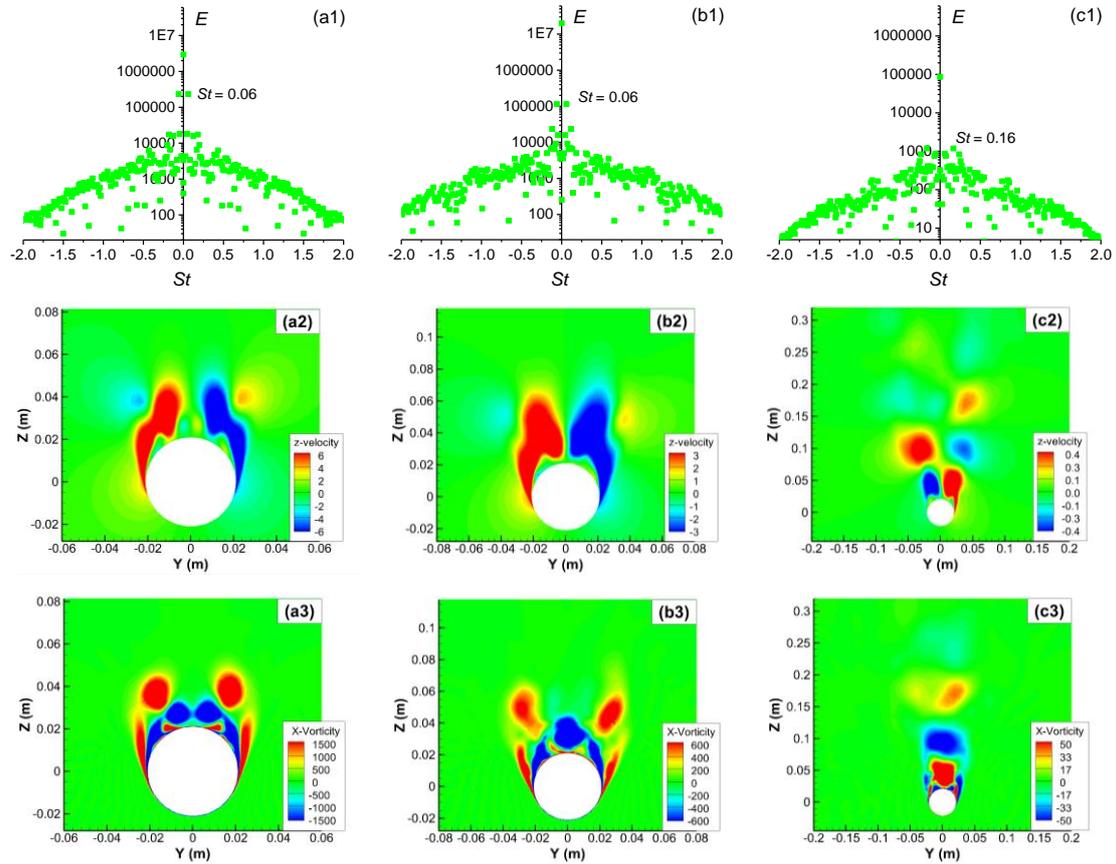

FIG. 10. Dynamic Mode Decomposition (DMD) at AOA=50º (z-velocity is z component velocity, m/s; x-Vorticity is x component of vorticity, 1/s) (a) x/D=2.5; (a) x/D=3.5; (c) x/D=16

The global modes of vortex oscillations depicted in (a2, a3) and (b2, b3) of Figure 9-11 correspond to alternate oscillations of the 3D vortex pairs, in which both the z-velocity and x-Vorticity can charaterize the mode patterns of this flow phenomenon well. Here the oscillatory global modes for vortex oscillations are all similar to the ones over a pointed-nose slender body[18]. These global modes are not studied extensively, partly because this phenomenon was found only in recent years, especially for the alternate oscillation of 3D vortices. By contrast, the vortex shedding is a classic phenomenon, and was investigated for a long time. The oscillatoty modes for vortex shedding are more common as shown in (c2, c3) of Figure 9-11. Nevertheless, it should be noted that for the case at lower AOAs, the downstream wakes of forebody vortices pass by over the afterbody, so the mode graphs at these sections of the afterbody will include the forebody vortex wakes, as shown in Figure 9 (c2, c3), where the top patterns as marked by a dash square are not part of the vortex shedding, and they come from the forebody wakes. This can be more clearly demonstrated by Figure 9 (c3), where this top pattern inside the dash square is apparently unlike the mode types of vortex shedding. Futhermore, it is well known that *St* of vortex shedding around a two dimensional cylinder at subcritical Reynolds number is about 0.21. However, for a three dimensional inclined cylinder, *St* will be less than 0.21 due to the three dimensionality and end effects [15]. At higher AOAs, the modes of vortex oscillations over the forebody sections are similar to the cases at lower AOAs, as shown in Figure 10 (a2-a3, b2-b3) and Figure 11 (a2-a3, b2-b3), but the modes of vortex shedding around the afterbody more and more approach the flow patterns of Kármán vortex street .

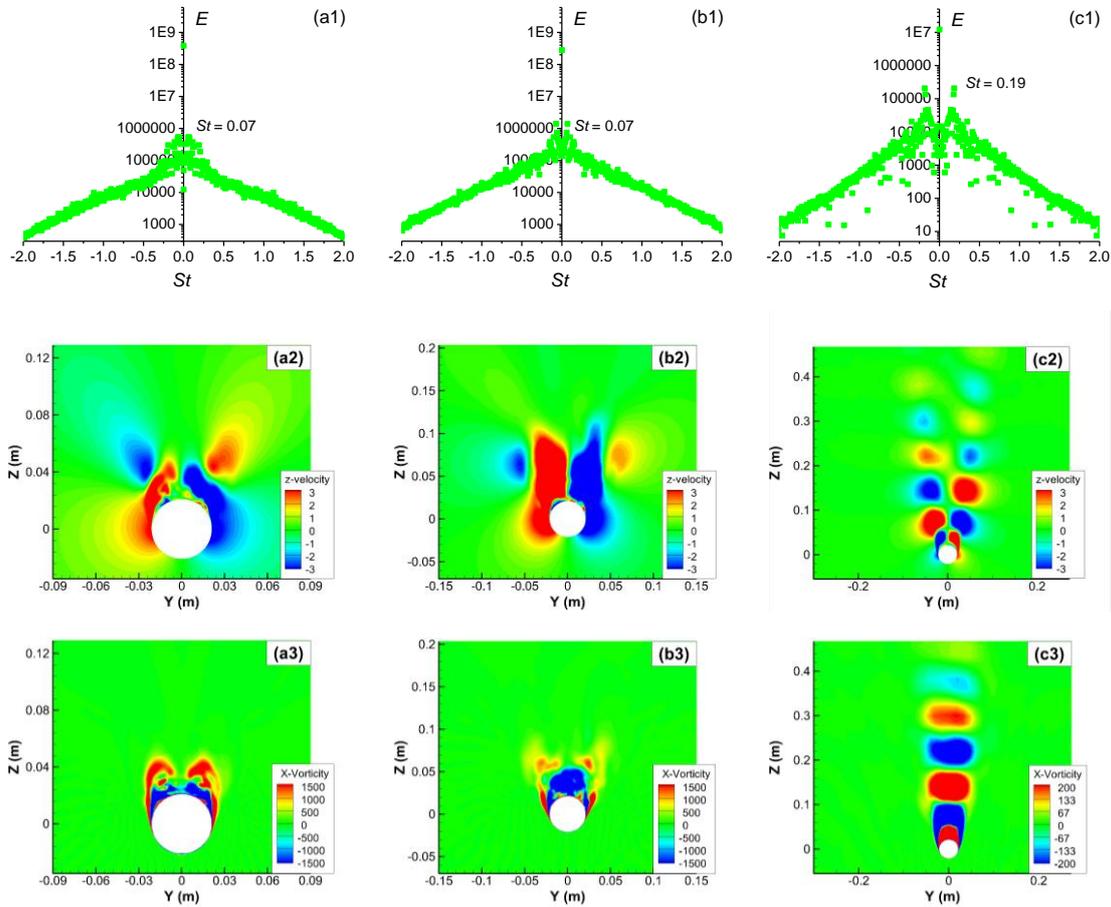

FIG. 11. Dynamic Mode Decomposition (DMD) at AOA=70º (z-velocity is z component velocity, m/s; x-Vorticity is x component of vorticity, 1/s)    (a) x/D=1.5; (b) x/D=2.5; (c) x/D=16

Figure 12 shows the distribution for the vortex oscillation and shedding on the map with the two parameters of AOAs and x/D. The boundary between the two phenomena clearly shows that there is a sharp forward movement for the vortex shedding as the AOAs less than 20º, then the regions of vortex shedding gradually expand with increasing AOAs.

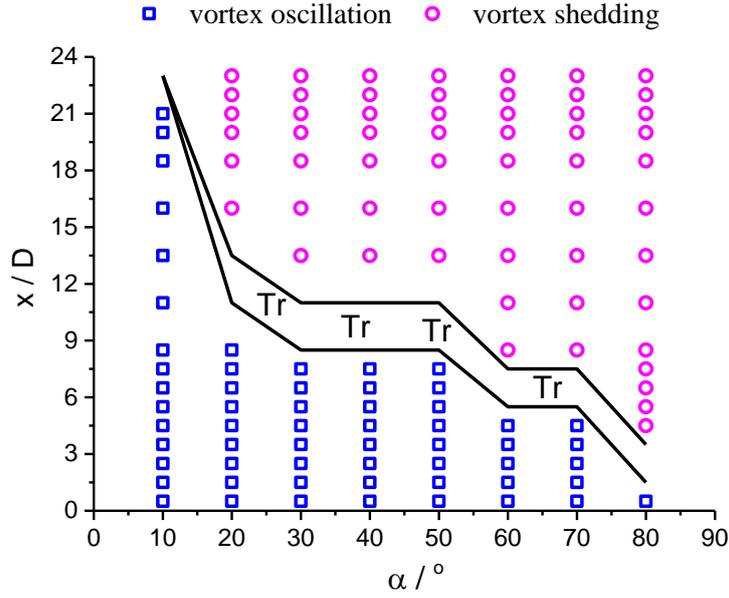

FIG. 12. Distribution map for vortex oscillations and shedding over a hemisphere-cylinder body at various AOAs and x/D, and *Tr* is the transitional region between vortex oscillations and shedding

## IV. DISCUSSION ON MECHANISMS OF VORTEX OSCILLATIONS

The above results present the unsteady flow fields, DMD modes and associated side forces over the hemisphere-nose slender body. These results reveal that the leeward vortex pair exhibits oscillation at lower AOAs (here 10°), and the vortex shedding occurs at higher AOAs and expand forward to the nose with AOAs increasing. The mechanisms governing the vortex shedding have been understood well in the previous studies on the flows around a circular cylinder, and basically the Von Kármán flow instability will be responsible for the formation of the vortex shedding. Of the vortex oscillation, however, the mechanisms are still not clear so far. Although the present investigation is probably unable to fully answer this question, some discussion based on the existing results is still interesting.

First of all, we need to determine whether the vortex oscillation is an independent unsteady phenomenon. Le Clainche et al. [29, 30] have ever studied the vortical flow around a hemisphere cylinder using a numerical simulation at Re=1000 and AOA=20°. In their results, the vortex shedding around the cylinder of the afterbody is not very strong due to using a model with a smaller fineness ratio of 8, but the vortex shedding still should exist at the rear end of the body, and the vortex pair on the leeward side of the body also exhibits oscillation. The authors suggested that it is difficult to conclusively state that the leeward vortex oscillation and unsteady afterbody wakes are independent phenomena or not, because they have the same oscillatory frequencies. This result is consistent with the cases at lower AOAs of 10°-40° in our present results where the frequencies along the body are also similar, so the mechanisms responsible for the two unsteady phenomena are really unable to be distinguished, as suggested by Le Clainche et al.[30]. However, the situation is different for higher AOAs of 40°-50° where the frequencies for vortex oscillations over the forebody and vortex shedding over the afterbody are significantly different, so the two types of unsteady phenomena should be independent at this range of AOAs. The vortex oscillation at this AOA range should be triggered by another vortex instability mechanism different from the one of vortex shedding, which is worthy of being investigated further. As for the cases of AOAs less than 40°, the frequencies are similar at the sections of the fore and after body, but not completely the same at most cases. More importantly, the behaviors of the vortex oscillation at the whole range of AOAs of 10°-80°, such as the vortex evolution with time and associated DMD modes, are much similar. Therefore, the mechanism for the vortex oscillation at these lower AOAs should be similar to the one at higher AOAs. Of course, the specific reason dominating the vortex oscillation need to be studied further in the future, and a global instability analysis [46] developed in recent years should be a good tool to reveal the mechanisms.

It is also interesting for comparing the flows over the hemisphere-nose and pointed-nose slender body. For unsteady behaviors of the flow fields, the biggest difference for both lie on the inception of the vortex oscillation. The forebody vortices over the hemisphere cylinder exhibit vortex oscillations at almost from low to high AOAs, but the vortex oscillation for pointed-nose bodies exists only at very high AOAs (generally more than 60º). One possible explanation for this difference is that the time-averaged vortex structure over the pointed-nose body is sensitive to tip irregularities and can become asymmetric at high AOAs. As a result, the asymmetric vortex pair becomes more stable owing to being far away from each other. Previous studies [18] have shown that the forebody vortices over the pointed-nose body develop downstream with straight streamwise vortices aligned basically with a freestream direction. However, here the downstream vortex wakes for the forebody vortices over the hemisphere cylinder body appear to lose stability, exhibiting helical vortex structures. It has been recognized that there are an elliptic instability and long-wave instability for a free vortex pair [45] which usually exist in the wakes after a large airplane, but both two instabilities have no oscillatory modes. Therefore, the instability and associated scenarios of 3D vortices over slender bodies probably have a new mechanism, which should be worth studying in the future.

## V. CONCLUDING REMARKS

The vortex unsteadiness around a hemisphere-cylinder body with a fineness ratio of 24 at the AOAs of 10º to 80º was studied using large eddy simulation and dynamic mode decomposition analysis. The results revealed that two types of unsteadiness exist over the hemisphere cylinder; one is vortex oscillations over the forebody, and another is vortex shedding over the downstream afterbody.

The two types of unsteady phenomena all induce fluctuating sectional side forces along the body, but the side forces induced by the vortex oscillation are larger and increase with increasing AOAs until around 60º AOA, after that, decrease. The frequencies of fluctuations of sectional side forces are between 0.085-0.12 at the AOAs of 10-40º where the frequencies of the vortex oscillation are similar to the vortex shedding at each AOA; while at the AOAs of 50-80º, the frequencies along the body are apparently divided into two regimes in which the vortex oscillation over the forebody have the frequencies of 0.053-0.064, and the vortex shedding over the afterbody has the frequencies of 0.16-0.2. Additionally, as the AOAs beyond 40º, the frequencies are basically proportional to the sine function of AOAs with the exceptions of the transitional region between the vortex oscillation and shedding as well as the region near the rear end of the body.

The vortex oscillation occurs even at the AOA of 10º, and exists within the whole range of AOAs, which is characterized by an alternate oscillation of a forebody leeward vortex pair in a direction normal to the freesteam and by associated in-phase swings from side to side. No vortex shedding occurs at the afterbody for the AOA of 10º, and the whole body exhibits a slight vortex oscillation, but the separation bubble near the hemisphere nose apparently shed downstream. The shedding of the separation bubble gradually disappears with AOAs increasing, which should contribute very little to the side forces owing to being far away from the body. The vortex shedding starts sharply from the afterbody at 20º AOA, as the forebody vortices break away from the body. As the AOAs less than 40º, the flow fields over the afterbody are dominated by vortex shedding and the downstream wakes of forebody vortices together. The forebody leeward vortex wakes twist together downstream, passing by over the shedding vortices at the afterbody. The region influenced by vortex shedding moves forwards gradually with AOAs increasing, and accordingly the region of vortex oscillations contracts and eventually only exists near the nose as AOAs sufficiently high.

The DMD analysis on the sectional flow fields shows that except for the mean flows, the most energetic modes correspond to the vortex oscillation at the forebody and to the vortex shedding at the afterbody, and the frequencies for the DMD modes are identical to the ones of the side forces obtained by Fourier transformation.

In addition, both time-averaged side forces and sectional vorticity fields show that the averaged vortex positions of the instantaneous vortex oscillation are symmetric with respect to the symmetry plane of the geometry, and no apparent asymmetry in position exists for the mean flow fields. Therefore, the leeward vortex pair over the forebody oscillates around a symmetric flow field.

**ACKNOWLEDGMENTS**

The project is supported by the National Natural Science Foundation of China under Grant Nos. 11272033.